\documentclass[12pt]{article}
\usepackage{graphicx}
\usepackage{hyperref}

\begin{document}

\hyphenpenalty=5000 \tolerance=1000

\begin{center}

{\Large \bf Do Small-mass Neutrinos participate in Gauge Transformations?}

\vspace{5mm}

Y. S. Kim\footnote{electronic mail: yskim@umd.edu} \\
Center for Fundamental Physics, University of Maryland, College Park,
MD 20742 USA

\vspace{5mm}

G .Q. Maguire Jr. \footnote{electronic mail: maguire@kth.se} \\
School of Information Technology, KTH Royal Institute of Techology, 
16440 Stockholm, Sweden \\

\vspace{5mm}

M. E. Noz \footnote{electronic mail: marilyn.noz@med.nyu.edu} \\
Department of Radiology, New York University, New York, NY
10016, USA \\

\end{center}

\vspace{5mm}

\begin{abstract}

Neutrino oscillation experiments presently suggest that neutrinos have
a small but finite mass. If neutrinos have mass, there should
be a Lorentz frame in which they can be brought to rest.  This paper
discusses how Wigner's little groups can be used to distinguish
between massive and massless particles.  We derive a representation of
the SL(2,c) group which separates out the two sets of spinors:
one set is gauge dependent and the other set is
gauge-invariant and represents polarized neutrinos. We show that a
similar calculation can be done for the Dirac equation.  In the
large-momentum/zero-mass limit, the Dirac spinors can be separated
into large and small components.  The large components are gauge
invariant, while the small components are not.  These small components
represent spin-$\frac{1}{2}$ non-zero mass particles. If we
renormalize the large components, these gauge invariant spinors
represent the polarization of neutrinos. Massive neutrinos cannot be
invariant under gauge transformations.

\end{abstract}
\vspace{30mm}

\newpage

\section{Introduction}\label{intro}

Whether or not neutrinos have mass and the consequences of this
relative to the Standard Model and lepton number are the subject of
much theorectical
speculation~\cite{papoulias_exotic_2013,dinh_observables_2013}, aa
well as
cosmological~\cite{miramonti_advancements_2013,higuera_measurement_2014,ahlers_high-energy_2015},
nuclear reactor~\cite{double_chooz_2014, li_unambiguous_2013}, and
high energy
experimentation~\cite{bergstrom_combining_2013,the_minos_collaboration_measurement_2013,han_lepton_2013,Aska2015}.
Neutrinos are fast becoming an important component of the search for
dark matter and dark
radiation~\cite{drewes_news_2013,ko_icecube_2015}.  Their importance
within the Standard Model is reflected in the fact that they are the
only particles which seem to exist with only one direction of
chirality, i.e., only left-handed neutrinos have been confirmed to
exist thus far. It was speculated some time ago that neutrinos in a
constant electric and magnetic fields would acquire a small mass, and
that right-handed neutrinos would be trapped within the interaction
field~\cite{barut_four_1986}. Additionally there are several physical
problems which right-handed neutrinos might help
solve~\cite{dre-gar-2015, Drewes_right-handed-2015,can-dre-gar-2014}. 
Solving generalized electroweak models using
left- and right-handed neutrinos has also been discussed~\cite{palcu_neutrino_2006}.  
Today right-handed
neutrinos which do not participate in weak interactions are called
``sterile'' neutrinos~\cite{bilenky_neutrino_2013}.  A comprehension
discussion of the place of neutrinos in the present scheme of particle
physics has been
given by Drewes~\cite{drewes_news_2013}.

In this paper we use representations of the Lorentz group to
understand the physical implications of neutrinos having mass.
In Sec.~\ref{lorentz-two}, two-by-two representations of the Lorentz
group are presented.  In Sec.~\ref{smm} the internal symmetries of
massive and massless particles are derived. A representation of the
SL(2,c) group, which separates out the two sets of spinors contained
therein, is presented in Sec.~\ref{sl2}. One set of spinors is gauge
dependent and represents massive particles. The other is gauge
invariant and represents polarized neutrinos. In Sec.~\ref{dirac}, we
show how in the large-momentum/zero-mass limit, the Dirac spinors can
be separated into two components, one of which can represent a
spin-$\frac{1}{2}$ non-zero mass particle. The question of gauge
invariance is then discussed. In Sec.~\ref{lorentz}, we discuss the
zero-mass limit and gauge invariance in the Lorentz transformation
framework.  Some concluding remarks are made in Sec.~\ref{conrm}.

\section{Representations of the Lorentz Group}\label{lorentz-two}

The Lorentz group starts with a group of four-by-four matrices
performing Lorentz transformations on the four-dimensional Minkowski
space of $(t, z, x, y)$ which leaves invariant the quantity $\left(t^2
- z^2 - x^2 - y^2 \right)$.  Since there are three generators of
rotations and three boost generators the Lorentz group is a
six-parameter group.

\par
Einstein observed that the Lorentz group is
also applicable to the four-dimensional energy and momentum
space of $\left(E, p_z, p_x, p_y \right).$  He
derived the Lorentz-covariant energy-momentum relation
commonly known as $E = mc^2$.  As this transformation leaves
$\left(E^2 - p_z^2 - p_x^2 - p_y^2 \right)$  invariant,  the particle
mass is a Lorentz invariant quantity. 

\par 
In his 1939 paper~\cite{wigner_unitary_1939}, Wigner studied the
symmetry properties of free particles by using operators which commute
with the specified four-momentum of the particle. His ``little
groups'' were defined to be those transformations that do
not change this four-momentum. For massive particles, the little
group is isomorphic to $O(3)$; indeed the $O(3)$-like little group's
kinematics is well understood. Massless particles are isomorphic
to the Euclidean group commonly known as E(2).  Wigner noted that the
E(2)-like subgroup of $SL(2,c)$ is isomorphic to the Lorentz group of
transformations~\cite{kimwigner87}, but the kinematics of this group
is not as well established as that of the $O(3)$-like little group as 
there is no Lorentz frame in which a
massless particle is at rest.

It is possible to construct the Lie
algebra of the Lorentz group from the three Pauli spin
matrices~\cite{dirac45b, naimark54, KimNoz86, bkn14sym}
as
\begin{equation}\label{lie05}
J_i = \frac{1}{2}\sigma_i, \quad\mbox{and}\quad
K_i = \frac{i}{2}\sigma_i ,
\end{equation}
These two-by-two matrices satisfy the 
following set of commutation relations:
\begin{equation}\label{lie01}
 \left[J_i, J_j\right] = i\epsilon_{ijk} J_k  \quad\quad
 \left[J_i, K_j\right] = i\epsilon_{ijk} K_k  , \quad\mbox{and}\quad
 \left[K_i, K_j\right] = -i\epsilon_{ijk} J_k  
\end{equation}
where the generators $J_i$ represent rotations and the generators
$K_i$ represent boosts.
There are six generators of the Lorentz group which
satisfy the three sets of commutation relations given in Eq.~(\ref{lie01}).
The Lie algebra of the Lorentz group consists of these
sets of commutation relations. 
\par
These commutation relations are invariant under Hermitian
conjugation, however, while the rotation generators are Hermitian, the
boost generators are anti-Hermitian:
\begin{equation}
  J_i^{\dag} = J_i, \quad\mbox{while}\quad  K_i^{\dag} = - K_i .
\end{equation}
Thus, it is possible to construct two representations of the Lorentz
group, one with $K_i$ and the other with $-K_i$.
For this purpose we shall use the notation~\cite{KimNoz86, Berestetskii82,bknIOP2015}
\begin{equation}
 \dot{K_i} = - K_i .
\end{equation}
\par
To demonstrate that this set of generators do 
perform Lorentz transformations, let us consider a point $X$ in four 
dimensional space such as the Minkowskian four-vector $(t, z, x, y)$. 
A Hermitian matrix of the form:
\begin{equation} \label{slc01}
X = \pmatrix{t + z & x - iy \cr x + iy & t - z} ,
\end{equation}
with determinant
\begin{equation}
   t^2 - z^2 - x^2 - y^2 ,
\end{equation}
can be written where all the components of $X$ are real. Indeed, every
Hermitian matrix can be written this way with real components.
Consider next a matrix of the form  
\begin{equation}\label{alphabeta}
 G = \pmatrix{\alpha & \beta \cr \gamma & \delta} ,
\end{equation}
with four complex matrix elements, thus eight real parameters, and
require that the determinant be equal to one.  If 
\begin{equation}\label{alphabeta2}
 G^{\dag} = \pmatrix{\alpha^{*} & \gamma^{*} \cr \beta^{*} & \delta^{*} },
\end{equation}
is the Hermitian conjugate of $G$, then 
\begin{equation}\label{slc02}
X^{\prime} = G~X~G^{\dag}
\end{equation}
defines a linear transformation
with real coefficents such that the
determinant of $X^{\prime}$ is equal to the determinant of $X$.  This constitutes
a real Lorentz transformation.  
The transformation of Eq.~(\ref{slc02}) can be explicitly written as
\begin{equation}\label{slc04}
\pmatrix{t^{\prime} + z^{\prime} & x^{\prime} - iy^{\prime} \cr x^{\prime} + iy^{\prime} & t^{\prime} - z^{\prime}} =
\pmatrix{\alpha & \beta \cr \gamma & \delta}
\pmatrix{t + z & x - iy \cr x + iy & t - z}
\pmatrix{\alpha^{*} & \gamma^{*} \cr \beta^{*} & \delta^{*} } .
\end{equation}
It is important to note that the transformation of Eq.~(\ref{slc02}) is not
a similarity transformation. In the $SL(2,c)$ regime, not all
the matrices are Hermitian~\cite{bkn14sym}.
Moreover, since the determinants of $G$
and $G^{\dag}$ are one, the determinant of $GG^{\dag}$ is also one.
As
\begin{equation}\label{alphabeta3}
Tr\left(GG^{\dag}\right) = (\alpha\alpha^{*} +
    \beta\beta^{*} +  \gamma\gamma^{*} +  \delta\delta^{*} ) \geq 1,
\end{equation}
Eq.(~\ref{slc02}) is a proper Lorentz transformation~\cite{bkn14sym,Bargman47,BK2006}.
\par
Since the determinant of $G$ is fixed and is equal to one, there are
six independent parameters. 
This six-parameter group is commonly called $SL(2,c)$.
As the Lorentz group has six generators, this two-by-two
matrix can serve as a representation of the Lorentz group.
\par
Likewise, the two-by-two matrix for the four-momentum of
the particle takes the form
\begin{equation}\label{mom22}
P = \pmatrix{p_0 + p_z &  p_x - ip_y \cr p_x + ip_y &  p_0 - p_z}
\end{equation}
with $p_0 = \sqrt{m^2 + p_z^2 + p_x^2 + p_2^2}.$
The transformation of this matrix takes the same form as that for
space-time given in Eqs.~(\ref{slc02}) and (\ref{slc04}).
The determinant of this matrix is $m^2$ and remains invariant under Lorentz
transformations. The explicit form of the transformation is
\begin{eqnarray}\label{trans15}
&{}& P^{\prime} = G~P~G^{\dag} =
\pmatrix{p^{\prime}_0 + p^{\prime}_z & p^{\prime}_x - ip^{\prime}_y \cr p^{\prime}_x + ip^{\prime}_y &  p^{\prime}_0 - p^{\prime}_z} \nonumber \\[2ex]
&{}&\hspace{5ex} = \pmatrix{\alpha & \beta \cr \gamma & \delta}
\pmatrix{p_0 + p_z &  p_x - ip_y \cr p_x + ip_y &  p_0 - p_z}
\pmatrix{\alpha^{*} & \gamma^{*} \cr \beta^{*} & \delta^{*}} .
\end{eqnarray}
It is this Lorentz invariant mass that is important for
discussing neutrino oscillation. In the next section the
internal symmetry of particles will be discussed using Wigner's little
groups.

\section{Internal Symmetries of Massive and Massless
  Particles}\label{smm}

When special relativity was formulated, the main focus was point
particles, without internal space-time
structures. How these
particles look to moving observers can be studied using 
Wigner's little groups~\cite{wigner_unitary_1939} where the 
subgroup of the Lorentz group whose
transformations leave the particle momentum invariant are considered.
However, the little groups can transform the internal
space-time structure of the particles.  Since the particle momentum is fixed
and remains invariant, it is possible to consider that the particle momentum is along the
$z$ direction.
\par
This momentum is thus invariant under rotations around the $z$ axis.  In addition,
these rotations commute with the Lorentz boost along the $z$
axis because, according 
to the Lie algebra of Eq.~(\ref{lie01}),
\begin{equation}
\left[J_3, K_3\right] = 0.
\end{equation}
\par
In Sec.~\ref{lorentz-two} it was shown that the Lorentz
transformation of the four-momentum
can be represented by two-by-two matrices and an explict form for this
transformation was given.
If the particle moves along the $z$ direction, the
four-momentum matrix becomes
\begin{equation}
P = \pmatrix{E + p & 0 \cr 0 & E - p},
\end{equation}
where $E$ and $p$ are the energy and the magnitude of momentum respectively.

\par
Let $W$ be a subset of matrices which leaves the
four-momentum invariant, then we can write
\begin{equation}\label{wigcondi31}
 P = W~P~W^{\dag} .
\end{equation}
These matrices constitute Wigner's little
groups dictating the internal space-time symmetry of the particle.

\par
If the particle is massive, it can be brought to the system where
it is at rest with $p = 0$. The four-momentum matrix is then
proportional to
\begin{equation}\label{massive}
         P = \pmatrix{1 & 0 \cr 0 & 1} .
\end{equation}
Since the momentum matrix is proportional
to the unit matrix, the $W$ matrix forms a unitary subset of the $G$
matrices and is Hermitian. The corresponding little
group is the $SU(2)$ subgroup of the Lorentz group.
It is sufficient to consider rotations around
the $y$ axis, as rotations around the $z$ axis do not change the
momentum. Thus the rotation matrix
\begin{equation}\label{3r01}
R(\theta) = \pmatrix{\cos(\theta/2) & -\sin(\theta/2) \cr
                  \sin(\theta/2) & \cos(\theta/2) }
\end{equation}
can be used. This forms 
a representation
of Wigner's $O(3)$-like little group for massive particles which
describes the spin orientation of the particle in the rest frame.

\par
For the massless particle, $E = p$.  Thus the four-momentum
matrix is proportional to
\begin{equation}\label{mzero}
         P = \pmatrix{1 & 0 \cr 0 & 0} ,
\end{equation}
and the Wigner matrix is
necessarily triangular and should take the form
\begin{equation}\label{tri31}
T(\gamma) = \pmatrix{1 & -\gamma \cr 0 & 1} .
\end{equation}
This matrix cannot be diagonalized.  Its inverse and Hermitian conjugate are
\begin{equation}\label{tri32}
T^{-}(\gamma) = \pmatrix{1 & \gamma \cr 0 & 1}, \quad\mbox{and}\quad
  T^{\dag}(\gamma) = \pmatrix{1 & 0 \cr -\gamma & 1} ,
\end{equation}
respectively.  Since the inverse is not the same as the Hermitian
conjugate, $T$ is not a Hermitian matrix. In order to preserve the
Lorentz properties of the boosted four momentum, $\gamma$ must be
real.  
\par 
To understand this better, consider that as the $O(3)$
group is contracted into the Euclidean group ($E(2)$) group, one can think
of $E(2)$ as a plane tangent to the North pole. Since $E(2)$ consists
of two translation operators and a rotation operator, the rotation
around the $z$ axis remains unchanged as the radius become large and
rotations around the $x$ and $y$ axes become translations in the $x$
and $-y$ directions respectively within the tangent plane.  For a
massless particle, the $E(2)$-like little group bears the same
relation to the $E(2)$ group as the $O(3)$-little group does to $O(3)$
for a massive particle. Thus Eq.~(\ref{tri31}) is the representation
of Wigner's $E(2)$-like little
group~\cite{kimwigner87,bknIOP2015,han_e2-like_1982} for massless
particles. It is now possible to apply this formalism to
spin-$\frac{1}{2}$ particles by considering the $SL(2,c)$ representation
of the Lorentz group.

\section{$SL(2,c)$ and Spinors}\label{sl2}

In the case of $SL(2,c)$, or
spin-$\frac{1}{2}$ particles, it is
necessary to consider both signs of the boost generators $K_i$. 
In Sec.~\ref{lorentz-two}, we
considered that
$SL(2,c)$ consists of non-singular two-by-two matrices which
have the form defined in Eq.~(\ref{alphabeta}).
This matrix is applicable
to spinors that have the form:
\begin{equation}\label{com4}
U = \pmatrix{1 \cr 0}, \quad\mbox{and}\quad V = \pmatrix{0 \cr 1} ,
\end{equation}
for spin-up and spin-down states respectively.
\par
Among the subgroups of $SL(2,c)$ there are $E(2)$-like little groups
which correspond to massless particles.  If we consider a massless
particle moving along the $z$ direction, then the little group is generated by $J_3,
N_1$, and $N_2$, where
\begin{equation}\label{com5}
N_1 = K_1 - J_2, \;\;\; N_2 = K_2 + J_1, \;\;\;  \mbox{and} \;\;\; J_3
= (1/2)\sigma_{3} .
\end{equation} 
As usual $J_3$ is the generator of rotations and the $N_i$ generate
translation-like transformations where:
\begin{equation}\label{trans1}
D(u,v) = D(u,0)D(0,v) = D(0,v) D(u,0)
\end{equation}
As these $N$ operators have been shown to be the generators of gauge transformations in the
case of the photon~\cite{kimwigner87}, they will be referred to as the gauge
transformation in the $SL(2,C)$
regime~\cite{wigner_unitary_1939,han_gauge-space_1982,han_space-time_1982,hks1986}.
Their role with respect to
massless particles of spin-$\frac{1}{2}$ will now be discussed~\cite{bkn14sym}.
\par
For massless spin-$\frac{1}{2}$
particles, the $J_i$ are still the
generators of rotations. However, because of the sign change allowed for
$K_i$ it is necessary to have two sets of $N_i$ operators designated as
$N_i^{(+)}$ and $N_i^{(-)}$, where, as defined in 
Eq.~(\ref{com5}), the $N_i^{(+)}$ have the explicit form
\begin{equation}\label{com6}
N_1^{(+)} = \pmatrix{ 0 & i\cr 0 & 0 }, \;\;
N_2^{(+)} =  \pmatrix{ 0 & 1 \cr 0 & 0 } .
\end{equation}
The Hermitian conjugates of the above $N_i^{(+)}$ provide $N_1^{(-)}$ and
$N_2^{(-)}$. Thus, there are two sets of boost generators involved.
\par
The transformation matrices defined in Eq.~(\ref{trans1}) can then be written as~\cite{HKSJMP1986}:
\begin{eqnarray}\label{com7}
D^{(+)}(u,v) = \exp(-i[uN_1^{[+]} + vN_2^{[+]}]) = \pmatrix{ 1 &
  u-iv \cr 0 & 1 } , \nonumber  \\
D^{(-)}(u,v) = \exp(-i[uN_1^{[-]} + vN_2^{[-]}]) = \pmatrix{ 1 &
  0 \cr -u-iv & 1 } .
\end{eqnarray}
Since there are two set of spinors in SL(2,c), the spinors whose
boosts are generated by $K_i = i/2\sigma$ will be written as  $\alpha$
for spin in the positive direction and $\beta$ for spin in the
negative direction. For the boosts generated by $K_i = -i/2\sigma$ 
we will use $\dot{\alpha}$ and $\dot{\beta}$. These spinors
are gauge-invariant in the sense that 
\begin{equation}\label{com8}
D^{(+)}(u,v)\alpha = \alpha, \;\; D^{(-)}(u,v)\dot{\beta} =
\dot{\beta}.
\end{equation}
However, if we carry out the explicit multiplication, these spinors
are gauge-dependent in the sense that  
\begin{equation}\label{com9}
D^{(+)}(u,v)\beta = \beta + (u-iv)\alpha, \;\; D^{(-)}(u,v)\dot{\alpha} =
\dot{\alpha} - (u + iv)\dot{\beta}
\end{equation}
The gauge invariant spinors of Eq.(\ref{com8}) appear as polarized
neutrinos~\cite{kimwigner87, han_e2-like_1982, kim_neutrino_2002}.
\par
Let us examine
further the gauge dependent
spinors of Eq.(\ref{com9}). To accomplish this, we construct unit
vectors in Minkowskian space by taking the direct product of two
SL(2,c) spinors:
\begin{eqnarray}\label{com10}  
-\alpha\dot{\alpha} = (1,i,0,0), \;\;  \beta\dot{\beta} = (1,-i,0,0) ,
\nonumber \\
\alpha\dot{\beta} = (0,0,1,1) ,  \;\; \beta\dot{\alpha} = (0,0,1,-1) .
\end{eqnarray}
This combines two half integer spins into integer spins. 
To make $D(u,v)$ consistent with Eq.(\ref{com10}), it is necessary to
choose
\begin{equation}\label{com11}
D(u,v) = D^{(+)}(u,v)D^{(-)}(u,v) ,
\end{equation}
where $D^{(+)}$ and $D^{(-)}$ apply to the first and second spinors of
Eq.(\ref{com10}) respectively.  Since the plane wave photon
four-potential does not depend on $\beta\dot{\alpha}$ because of the
Lorentz condition~\cite{kimwigner87,han_e2-like_1982,
  han_gauge-space_1982,HKSJMP1986,kim_neutrino_2002}, 
we have  
\begin{eqnarray}\label{com12} 
D(u,v)(-\alpha\dot{\alpha}) = -\alpha\dot{\alpha} +
(u+iv)\alpha\dot{\beta}, \nonumber \\
D(u,v)(\beta\dot{\beta}) = \beta\dot{\beta} +
(u-iv)\alpha\dot{\beta}, \nonumber \\
D(u,v)\alpha\dot{\beta} = \alpha\dot{\beta}.
\end{eqnarray}
The first two equations in Eq.(\ref{com12}) correspond to gauge
transformations of the photon polarization vectors.  The third
equation corresponds to the effect of the $D$ transformation of the
four-momentum.  This shows that $D(u,v)$ is an element of the little
group. We look next at how we can apply this analysis to Dirac
spinors.  

\section{Dirac Spinors and Massless Particles}\label{dirac}

The Dirac equation is applicable to massive particles.  Here we will
consider the massless particle as the limiting case of the massive
particle by considering the large-momentum/zero-mass limit of the Dirac
spinors.  
\par
Starting with the spin operators defined in Eq.(\ref{lie05}), a boost
along the $z$ direction will take the form:
\begin{equation}\label{mas1}  
J^{\prime}_i = B(P)J_iB^{-1}(P).
\end{equation}
This is a similarity transformation.
Here the boost matrix is given by:
\begin{equation}\label{mas2}
B(P) = \pmatrix{e^{\eta /2} & 0 \cr 0 & e^{-\eta /2} },
\end{equation}
where 
\begin{equation}\label{mas2a}
e^{\eta/2} = \left ( \frac{E+P}{E-P} \right ).
\end{equation}
In the large-momentum or large-mass limit for a massive particle we obtain:
\begin{equation}\label{mas3}
e^{\eta} \rightarrow \frac{2E}{M}.
\end{equation}
Using the similarity transformation of Eq.(\ref{mas1}),
$J_3$ is invariant, but $J_1$ and $J_2$ take the form:
\begin{equation}\label{mas4}
J_1^{\prime} = \pmatrix{ 0 & \frac{1}{2}e^{\eta} \cr
  \frac{1}{2}e^{-\eta} & 0} \;\;\;\;
J_2^{\prime} = \pmatrix{ 0 & - \frac{i}{2}e^{\eta} \cr
  \frac{i}{2}e^{-\eta} & 0} .
\end{equation}
In the large-momentum or large-mass limit for a massive particle, we can obtain the $N_i$
matrices of Eq.(\ref{com5}) as:
\begin{equation}\label{mas5}
N_1 = -\frac{M}{E} J_2^{\prime}  \;\;\;\;  N_2 = \frac{E}{M}
J_1^{\prime} .
\end{equation}
\par
Remembering that we have to consider both signs of the boost
generators, the generators of $SL(2,c)$ can take the form:
\begin{equation}\label{mas6}
J_i = \pmatrix{(1/2)\sigma_i & 0 \cr 0 & (1/2)\sigma_i } , \;\;\;\;
K_i = \pmatrix{(i/2)\sigma_i & 0 \cr 0 & (-i/2)\sigma_i } 
\end{equation}
which is applicable to Dirac wave functions in the Weyl
representation~\cite{barut_four_1986,KimNoz86}. Using the gauge transformation
matrices from Eq.(\ref{com7}), we can write:
\begin{equation}\label{mas7}
D(u,v) = \pmatrix{D^{(+)} (u,v) & 0 \cr 0 &  D^{(-)} (u,v) }.
\end{equation}
This matrix is applicable to the Dirac spinors.  To evaluate the
result of applying the $D$ matrix from Eq.(\ref{mas7}), we first look
at the eigenspinors given in Eq.(\ref{com4}) applied to a massive Dirac
particle that is at rest.   Thus we have:
\begin{equation}\label{mas8}
U(0) = \pmatrix{\alpha \cr \pm\dot\alpha} \;\;\;\; V(0) = \pmatrix{\pm\beta \cr \dot\beta},
\end{equation}
where the positive and negative energy states are denoted by the $+$
and $-$ signs respectively. If these spinors are boosted along the
z-axis using the operator generated by $K_3$, then 
\begin{equation}\label{mas9}
U(P) = \pmatrix{e^{(+\eta/2)}\alpha \cr
  \pm e{^{(-\eta/2)}}\dot\alpha} ,  \;\;\;\; V(P) =
\pmatrix{\pm e{^{(-\eta/2)}}\beta \cr e{^{(+\eta/2)}}\dot\beta}.
\end{equation}
In the large-momentum/zero-mass limit, the large components,
$e^{(+\eta/2)}$, are, according to Eq.(\ref{com8}),
gauge invariant, while the small components acording to Eq.(\ref{com9}), are
gauge dependent.  This again shows that non-zero mass, spin-$\frac{1}{2}$
particles are not invariant under gauge transformations. Furthermore,
in this limit, the spinors of Eq.(\ref{mas9}) can be renormalized as:
\begin{equation}\label{mas10}
U(P) = \pmatrix{\alpha \cr 0 } \;\;\;\; V(P) = \pmatrix{0 \cr \dot\beta},
\end{equation}
It is clear that the $D$ transformation leaves these spinors
invariant.  It is this invariance, as shown before, that is responsible for the
polarization of neutrinos~\cite{han_e2-like_1982,kim_neutrino_2002}. 

Additionally, one could interprete the results of  Eq.(\ref{mas9}) in
terms of $E(2)$ translations on free Weyl neutrino states. In this case,
the gauge invariant transformations leave the left-handed neutrino
invariant, but translate the right-handed neutrino into a linear
combination of left-handed and right-handed
neutrinos~\cite{barut_four_1986,han_e2-like_1982}. These coupled
states could have implications requiring that in a constant electric
and magnetic field neutrinos should acquire a small effective
mass~\cite{barut_four_1986}.

\section{Neutrino Mass and Lorentz Transformations}\label{lorentz}

In Sec.~\ref{smm}, we introduced the fact that
Wigner~\cite{wigner_unitary_1939} proposed 
that his ``little groups'' be defined as those 
Lorentz transformations that do not change the four-momentum of the free particle.
Because there is no Lorentz frame in which a massless particle is at rest, we had
to consider a momentum four-vector of the form given in
Eq.(\ref{mzero}). From this we were able to write down the
transformation matrix given in Eq.(\ref{tri31}) which left the
four-momentum invariant.  In this section, we begin with 
a massive particle with fixed energy $E$.
Then we can write the Lorentz boost along the $z$ direction as
\begin{equation} \label{eq301}
   z \rightarrow (\cosh\xi) z + (\sinh\xi) t, \qquad
   t \rightarrow (\sinh\xi) z + (\cosh\xi) t.
\end{equation}
As we saw in Sec.~\ref{dirac} the limiting case of $e^{\xi}$ is given
in Eq.(\ref{mas3}) for the large momentum limit. 
Within the framework of Lorentz transformations, $E$ can become large and thus
$\xi$ can also become large. 
This has been discussed in the
literature~\cite{KimNoz86}.
\par
In addition, $\xi$ can become large when the mass becomes
very small.  This cannot be achieved by Lorentz boosts, because the
mass is a Lorentz-invariant quantity.  With this point in mind, we
can consider what happens when the mass is varied but the energy is
held fixed. We can
write the energy-momentum four-vector as
\begin{equation}\label{eq303}
E(0, 0, \cos\chi, 1) .
\end{equation}
Then the mass becomes
\begin{equation}\label{eq304}
M = E \sqrt{1 - \cos^2\chi} = E \sin\chi .
\end{equation}
Hence, the mass can be increased by increasing $\chi$ from $zero$~\cite{kim_noz_2013}.
\par
While the four-by-four matrix which makes the transformation of Eq.(\ref{eq301})
is
\begin{equation}\label{eq305}
\pmatrix{1 & 0  & 0 & 0 \cr
0 & 1 & 0 & 0 \cr 0 & 0 & \cosh\xi & \sinh\xi \cr 0 & 0 & \sinh\xi &
\cosh\xi } ,
\end{equation}
its two-by-two equivalent to the spinor is
\begin{equation}\label{eq306}
\pmatrix{e^{\xi/2} & 0 \cr 0 & e^{-\xi/2} }
\end{equation}
as seen in Eq.(\ref{mas2})~\cite{kim_noz_2013}.
The two-by-two matrix corresponding to a rotation around the $y$ axis is
\begin{equation}\label{eq307}
\pmatrix{\cos(\theta/2) & -\sin(\theta/2) \cr \sin(\theta/2) & \cos(\theta/2)} .
\end{equation}
Thus we can now perform the Lorentz boost by making a similarity transformation:
\begin{equation}\label{eq308}
\pmatrix{e^{\xi/2} & 0 \cr 0 & e^{-\xi/2} }
   \pmatrix{\cos(\theta/2) & -\sin(\theta/2) \cr \sin(\theta/2) & \cos(\theta/2)}
\pmatrix{e^{-\xi/2} & 0 \cr 0 & e^{\xi/2} },
\end{equation}
which becomes
\begin{equation}\label{eq309}
   \pmatrix{\cos(\theta/2) & - e^{\xi} \sin(\theta/2) \cr
        e^{-\xi}\sin(\theta/2) & \cos(\theta/2)} .
\end{equation}
For this matrix to remain finite in the large $\xi$ limit, we can 
let $e^{\xi}\sin\theta = \gamma$. For $\gamma$ to remain finite as
$\xi$ increases, $\theta$ must approach zero. Then, in the limiting case, the matrix given
in Eq.(\ref{eq309}) becomes
\begin{equation} \label{eq310}
\pmatrix{1 & -\gamma \cr 0 & 1} .
\end{equation}
It has been shown that the $\gamma$ parameter performs gauge transformations
on the photon case, and its equivalent transformation on massless
neutrinos~\cite{KimNoz86,han_e2-like_1982,kim_neutrino_2002}.  If the neutrino
indeed has mass, then we should observe 
neutrinos participating in gauge
transformations.

\section{Concluding Remarks}\label{conrm}
As there is currently much interest in massive neutrinos, it would be
interesting to see if there was indeed a Lorentz frame in which
neutrinos could be brought to rest.  Additionally, it would useful to
understand if neutrinos participate in gauge transformations.
The issue of whether or not the neutrino is a Dirac particle as
opposed to a Majorana particle will be settled only if lepton number
violation is observed.  Furthermore, if right-handed neutrinos could
be found separated from left-handed neutrinos, and if these
right-handed neutrinos did not partcipate in weak interactions, this would have
implications for physics beyond the Standard Model.


\end{document}